# PRE-SOLUTION OF THE PERTURBED MOTION OF ARTIFICIAL SATELLITE

*I.A. HASSAN[1], Z.M. HAYMAN[2] and M.A.F. Basha[3]*

[1]*Dept. of Astronomy, Fac. Sci., AL-Azhar University, Egypt.*

[2]*Dept. of Astronomy, Fac. Sci., Cairo University, Egypt,*

[3]*Dept. of Physics, Fac. Sci., Cairo University, Egypt.*

**Abstract**

The authors try to find a good solution of an artificial satellite motion under the influence of $J_2$–gravity in terms of KS variables by using Picard's Iterative Method. The result shows that there are many solutions for this problem depends on the initial guess solutions, so the choice of correct/convince initial guess is very difficult. Applications of the method applied on many satellites.

**1. Introduction**

It is well known that the solutions of the Classical Newtonian Equations of motion are unstable and these equations are not suitable for long-term integration. So, many transformations have emerged in the literature in the recent past, for example the set of KS variables (Stiefel and Scheifele, 1971). So, the yield equations are the equations of motion in terms of the KS variables include perturbations that can arise from a potential and perturbations that cannot be derived from it, also they are valid for any type of orbital motion.

Accurate orbit prediction of the Earth's satellites is an important problem for mission planning, coverage, geodesy decay and lifetime estimates, so some difficulties are facing the conventional numerical methods used in Celestial Mechanics such as the urgent need for a narrow step length, instability, the sensitivity to round-off and truncation errors, besides the collision or close approach difficulties. Such difficulties initiated and activated the search of different sets of the so-called regularizing variables that do not suffer such inconveniences. These regularized variables though proved to be useful in numerical applications, are still not common in analytical work.

In this work, we introduce pre-solutions of the perturbed differential equations of any satellite in terms of KS variables depending on Picard's iterative method (Piaggio, 1958; Struble, 1962; Beyn, 1990 and Parker et al., 1996).

## 2. Formulation of the Problem:

The equations of motion of an artificial satellite are given generally as

$$\ddot{\vec{x}} + \frac{\mu}{r^3}\vec{x} = -\frac{\partial V}{\partial \vec{x}} + \vec{P} \qquad (2.1)$$

where $\vec{x}$ is the position vector in a rectangular frame (the physical frame), $r=|\vec{x}|$ is the distance from the origin, $\mu$ is the Earth's gravitational constant, V is the perturbed time independent potential and $\vec{P}$ is the resultant of all non-conservative perturbing forces and forces derivable from a time dependent potential.

The potential of the Earth's gravity with axial symmetry can be written as

$$V = \mu \sum_{i=2}^{\infty} R^i J_i \left(\frac{1}{r}\right)^{i+1} P_i(x_3/r), \qquad (2.2)$$

where $R$ is the Earth's equatorial radius, $J_i$ is the non-dimensional coefficient of the Earth's oblateness and $P_i(x_3/r)$ is the Legendre polynomial of order $i$. In the present paper we'll take the effect of the Earth's gravity oblateness, then Eq.(2.2) rewrite as

$$V = \frac{3}{2} Q_2 \, x_3^2 \, r^{-5} - \frac{1}{2} Q_2 \, r^{-3}. \qquad (2.3)$$

where $\qquad Q_2 = \mu R^2 J_2$,

and $\qquad r = \sqrt{x_1^2 + x_2^2 + x_3^2}$.

Finally, the equations of motion of an artificial satellite in KS-regularized variables are

$$\vec{u}'' + \alpha_k \vec{u} = \frac{r}{2} \vec{\lambda}, \qquad (2.4.1)$$

$$\alpha_k = -\langle \vec{u}', \vec{\lambda} \rangle, \qquad (2.4.2)$$

$$t' = r, \qquad (2.4.3)$$

$$r'' + 4\alpha_k r = \mu + r \langle \vec{u}, \vec{\lambda} \rangle, \qquad (2.4.4)$$

where

- $\alpha_k$ is one-half of the negative Keplerian energy

$$\alpha_k = \left(\frac{\mu}{2} - \langle \vec{u}', \vec{u}' \rangle\right)/r \;;$$

- $\vec{\lambda} = L^T(\vec{u}) \vec{b} = L^T(\vec{u})\left(-\frac{\partial V}{\partial \vec{x}} + \vec{P}\right),$

- $L(\vec{u}) = \begin{pmatrix} u_1 & -u_2 & -u_3 & u_4 \\ u_2 & u_1 & -u_4 & -u_3 \\ u_3 & u_4 & u_1 & u_2 \\ u_4 & -u_3 & u_2 & -u_1 \end{pmatrix},$

- $r = -<\vec{u},\vec{u}>$, and $r' = 2<\vec{u},\vec{u}'>$;

hence $<\vec{a},\vec{b}>$ is used to denote the scalar product of two vectors $\vec{a}$ and $\vec{b}$. Denoting differentiation with respect to the new time s (known as the fictitious time) by a prime ('), since the independent variable is changed from time (t) to fictitious time (s) according to (Stiefel and Scheifele, 1971)

$$\frac{dt}{ds} = r.$$

## 3. Equations of Motion:

The differential equations of motion for the satellite in KS-regularized variables under the perturbations of the Earth's gravity and air drag are

$$u_1'' = -\alpha_k u_1 + \frac{r}{2}\lambda_1, \tag{3.1}$$

$$u_2'' = -\alpha_k u_2 + \frac{r}{2}\lambda_2, \tag{3.2}$$

$$u_3'' = -\alpha_k u_3 + \frac{r}{2}\lambda_3, \tag{3.3}$$

$$u_4'' = -\alpha_k u_4 + \frac{r}{2}\lambda_4, \tag{3.4}$$

$$\alpha_k' = -u_1'\lambda_1 - u_2'\lambda_2 - u_3'\lambda_3 - u_4'\lambda_4, \tag{3.5}$$

$$t' = r, \tag{3.6}$$

$$r'' = \mu + r\left(-4\alpha_k + u_1\lambda_1 + u_2\lambda_2 + u_3\lambda_3 + u_4\lambda_4\right), \tag{3.7}$$

where

$$\lambda_1 = u_1 b_1 + u_2 b_2 + u_3 b_3,$$
$$\lambda_2 = -u_2 b_1 + u_1 b_2 + u_4 b_3,$$
$$\lambda_3 = -u_3 b_1 - u_4 b_2 + u_1 b_3,$$
$$\lambda_4 = u_4 b_1 - u_3 b_2 + u_3 b_3;$$

and

$$b_1 = \frac{15}{2} Q_2 x_1 x_3^2 r^{-4} - \frac{3}{2} Q_2 x_1 r^{-5},$$

$$b_2 = \frac{15}{2} Q_2 x_2 x_3^2 r^{-4} - \frac{3}{2} Q_2 x_2 r^{-5},$$

$$b_3 = -\frac{9}{2} Q_2 x_3 r^{-5} + \frac{15}{2} Q_2 x_3^3 r^{-7}.$$

## 4. Picard's Iterative Method:

Indeed, often it is very hard to solve differential equations, but do have a numerical process that can approximate the solution. This process is known as the Picard iterative process.

Picard was among the first to look at the associated functional equation. His method of finding the solution is known as the method of successive approximations or Picard's iteration method.

The Picard iterative process consists of constructing a sequence of functions that will get closer and closer to the desired solution.

Now, let us introduce this method. If we start with the general, first order nonlinear differential equation

$$\frac{dy}{dx} = f(x, y),$$

with the initial condition

$$y(x_0) = y_0,$$

we can integrate both sides of the equation for $x_0 \leq x \leq x_1$ to get

$$y(x_1) - y(x_0) = \int_{x_0}^{x_1} dy = \int_{x_0}^{x_1} f(x, y(x)) \, dx.$$

Using $t$ as the variable for integration, instead of $x$, and setting $x = x_1$ to be the arbitrary final point

$$y(x) = y(x_0) + \int_{x_0}^{x} f(t, y(t)) \, dt$$

We proceed to try to solve this by iteration. Substituting an *initial guess* of $y(x) = \phi_0(x)$ into the right hand side of the above equation, we get

$$\phi_1(x) = y(x_0) + \int_{x_0}^{x} f(t, \phi_0(t)) \, dt,$$

where $\phi_1(x)$ is the approximate solution.

In the unlikely event that $\phi_1(x) = \phi_0(x)$, we are finished, we have found a solution. Otherwise, we keep on iterating

$$\phi_{n+1}(x) = y(x_0) + \int_{x_0}^{x} f(t, \phi_n(t)) \, dt.$$

If no further information on the solution is available, we normally choose $\phi_0(x) = y_0$.

Generally,

$$\phi_{n+1}^{(k)}(x) = y^{(k)}(x_0) + \int_{x_0}^{x} f(t, \phi_n^{(k)}(t)) \, dt,$$

where $k$ (= 1, 2, ..., n) is the number of n variables.

**5. The Methodology:**

In this section, Picard's iterative solution of Eqs.(3.1) to (3.7) will be gone in the two following steps. The <u>first step</u> is rewriting the equations of motion in terms of independent variables (*u's*) only, and the <u>second stage</u> is constructing the algorithm to solve the resultant equations by using Picard's iterative method.

*Step1:*

The first step is to transform Eqs.(3.1) to (3.7) into first order differential equations by the following substitutions

$$y_i = u_i, \qquad y_{i+4} = u'_i, \quad i = 1(1)4,$$
$$y_9 = \alpha_k, \qquad y_{10} = t, \qquad y_{11} = r \quad \text{and} \quad y_{12} = r'.$$

Then the first order system of the problem becomes

$$y'_1 = y_5, \tag{4.1.1}$$
$$y'_2 = y_6, \tag{4.1.2}$$
$$y'_3 = y_7, \tag{4.1.3}$$
$$y'_4 = y_8, \tag{4.1.4}$$
$$y'_5 = -y_9 y_1 + \tfrac{1}{2} y_{11} b_1, \tag{4.1.5}$$
$$y'_6 = -y_9 y_2 + \tfrac{1}{2} y_{11} b_2, \tag{4.1.6}$$
$$y'_7 = -y_9 y_3 + \tfrac{1}{2} y_{11} b_3, \tag{4.1.7}$$
$$y'_8 = -y_9 y_4 + \tfrac{1}{2} y_{11} b_4, \tag{4.1.8}$$
$$y'_9 = -y_5 b_1 - y_6 b_2 - y_7 b_3 - y_8 b_4, \tag{4.1.9}$$
$$y'_{10} = y_{11}, \tag{4.1.10}$$
$$y'_{11} = y_{12}, \tag{4.1.11}$$
$$y'_{12} = \mu + y_{11} \left( y_1 b_1 + y_2 b_2 + y_3 b_3 + y_4 b_4 - 4 y_9 \right). \tag{4.1.12}$$

*Step 2:* The algorithm is
1- Set the Eqs.(4.1).
2- Set the initial guess of the solutions of the equations, i.e.,
$$y_i = \text{Initial guess.}$$
3- Get the pre-solutions by integration the equations, as
$$y_i = y_i(0) + ExpandAll[Integrate[y_i, \phi_o, \phi]], \quad i = 1(1)12.$$
4- Substitute the pre-solutions in the original Eqs.(4.1) to get the new one which will be integrated again.

## 6. Results and Conclusion:

We have two results of this paper. The first result is getting the equations of motion of an artificial satellite under the influence of Earth's oblateness in terms of KS variables with independent variables ($y$'s). These equations can be solved by any methods. The second result is used Picard's iterative method to get the pre-solutions, which listed in computational algorithm. The final iterative solutions of this method depend on the initial guess of the solutions. Therefore, the authors suffer to get the correct solutions or the pre-solution. So, this method of solution is not appropriate to find a solution of perturbed motion of an artificial satellite in terms of KS variables.